\documentclass[conference, 10 pt]{IEEEtran}

\usepackage{soul} 
\IEEEoverridecommandlockouts
\usepackage{cite}
\usepackage{amsmath,amssymb,amsfonts}
\usepackage{braket}
\usepackage{algorithmic}
\usepackage{multirow}
\usepackage{graphicx}
\usepackage{textcomp}
\usepackage{xcolor}
\def\BibTeX{{\rm B\kern-.05em{\sc i\kern-.025em b}\kern-.08em
    T\kern-.1667em\lower.7ex\hbox{E}\kern-.125emX}}
\begin{document}

\title{TrojanNet: Detecting Trojans in Quantum Circuits using Machine Learning\\
}

\author{\IEEEauthorblockN{Subrata Das}
\IEEEauthorblockA{\textit{Dept. of Electrical Engineering} \\
\textit{Pennsylvania State University}\\
University Park, PA \\
sjd6366@psu.edu}
\and
\IEEEauthorblockN{Swaroop Ghosh}
\IEEEauthorblockA{\textit{Dept. of Electrical Engineering} \\
\textit{Pennsylvania State University}\\
University Park, PA \\
szg212@psu.edu}

}

\maketitle

\begin{abstract}
Quantum computing holds tremendous potential for various applications, but its security remains a crucial concern. Quantum circuits need high-quality compilers to optimize the depth and gate count to boost the success probability on current noisy quantum computers. There is a rise of efficient but unreliable/untrusted compilers; however, they present a risk of tampering such as Trojan insertion. We propose TrojanNet, a novel approach to enhance the security of quantum circuits by detecting and classifying Trojan-inserted circuits. In particular, we focus on the Quantum Approximate Optimization Algorithm (QAOA) circuit that is popular in solving a wide range of optimization problems. We investigate the impact of Trojan insertion on QAOA circuits and develop a Convolutional Neural Network (CNN) model, referred to as TrojanNet, to identify their presence accurately. Using the Qiskit framework, we generate 12 diverse datasets by introducing variations in Trojan gate types, the number of gates, insertion locations, and compiler backends. These datasets consist of both original Trojan-free QAOA circuits and their corresponding Trojan-inserted counterparts. The generated datasets are then utilized for training and evaluating the TrojanNet model. Experimental results showcase 
an average accuracy of 98.80\% and an average F1-score of 98.53\% in effectively detecting and classifying Trojan-inserted QAOA circuits. Finally, we conduct a performance comparison between TrojanNet and existing machine learning-based Trojan detection methods specifically designed for conventional netlists.



\end{abstract}


\section{Introduction}
\label{sec:intro}
Quantum computing has emerged as a groundbreaking technology with the potential to revolutionize various fields such as optimization, cryptography, and material science \cite{national2019quantum}. Quantum Approximate Optimization Algorithm (QAOA) circuits play a crucial role in solving a wide range of combinatorial optimization problems using quantum computers \cite{zhou2020quantum, alam2020accelerating}. These circuits enable the exploration of the optimization landscape 
using variational techniques to find near-optimal solutions, making them invaluable tools for addressing complex optimization tasks. Their versatility and scalability have garnered significant interest, positioning QAOA circuits as a fundamental component of quantum computing research and applications. 
These circuits often encapsulate proprietary algorithms, novel optimization strategies, or domain-specific knowledge. For instance, a QAOA circuit designed for optimizing financial portfolio management may encapsulate an intricate combination of portfolio risk assessment and asset allocation techniques.

In order to fully leverage the promising advantages of quantum computing, it is of utmost importance to prioritize its security and privacy \cite{upadhyay2023obfuscating, das2023randomized}. Quantum circuits, especially those utilizing QAOA, heavily rely on high-quality compilers to optimize their performance \cite{shaydulin2019hybrid}. These compilers play a crucial role in reducing the depth and gate count of quantum circuits, which in turn increases the likelihood of success on quantum computers that are susceptible to noise and errors. However, recent developments have revealed the emergence of numerous efficient third-party compilers that claim to offer superior optimization capabilities for complex quantum circuits compared to their well-established counterparts \cite{zapatacomputing_2023, pytket}. While these compilers may seem attractive in terms of efficiency, their reliability is uncertain. One of the main risks associated with relying on unreliable compilers is the potential for tampering, specifically the insertion of Trojan gates. These unverified compilers may lack proper security safeguards or fail to adhere to trusted protocols
posing a serious threat to the integrity and trustworthiness of quantum computing systems.

\begin{figure}[t]
    \includegraphics[width=0.5\textwidth]{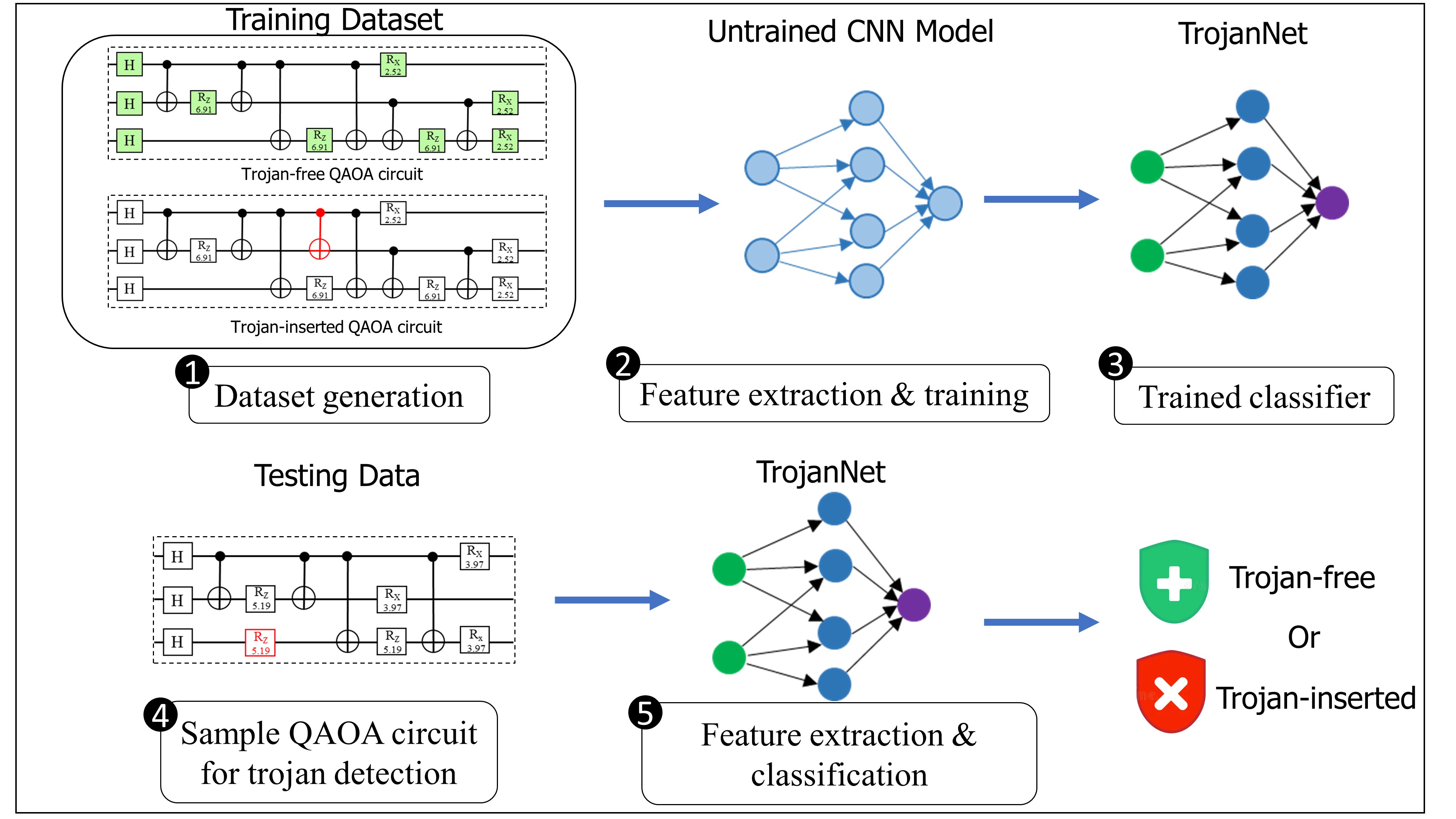}
    \centering
\caption{Overview of the proposed machine learning based Trojan detection technique.}
    \label{fig:schematic}
    \vspace{-.31cm}
\end{figure}

\textbf{Proposed Idea:}
\label{subsec:proposed_idea} 
To enhance the security of QAOA circuits and detect Trojan-inserted circuits, we propose TrojanNet that leverages Convolutional Neural Networks (CNNs). 
First, we explore various strategies to insert Trojans or extra gates, their types and insertion locations into QAOA circuits to maximize the
impact on the optimization process. Then, we generate 12 different datasets of compiled Trojan-free and Trojan-inserted QAOA circuits by varying Trojan-type, numbers, insertion locations and compiler backends. The generated datasets are used to train the TrojanNet CNN model. This model is specifically tailored to capture the subtle patterns and features in QAOA circuits that indicate the presence of Trojans. For instance, a typical QAOA circuit for the Max-Cut problem follows a pattern of alternating mixing and cost operator layers. The mixing layer consists of R\textsubscript{x} rotation gates, while the cost layer includes Controlled-X (CX) gates and R\textsubscript{z} rotation gates. Trojans disrupt this pattern by inserting extra gates and modifying the circuit structure, thereby breaking the regularity that QAOA circuits exhibit. TrojanNet can effectively detect Trojans by identifying these disruptions in the circuit's pattern. Fig. \ref{fig:schematic} illustrates this idea further.



\textbf{Contributions: }\label{subsec:our_contributions} 
Firstly, we propose a strategy for inserting Trojans or extra gates in QAOA circuits, considering different gate types and insertion locations. This comprehensive exploration allows us to identify the configurations that have the most significant impact on the optimization process. By quantitatively assessing the optimization results, we provide insights into the most critical areas of vulnerability and potential attack vectors. 
Secondly, we generate a comprehensive dataset comprising original Trojan-free QAOA circuits and their corresponding Trojan-inserted counterparts. Thirdly, we introduce TrojanNet to detect and classify Trojan-inserted QAOA circuits. TrojanNet leverages the power of deep learning to capture subtle patterns and features within the circuits, enabling accurate identification of Trojans. Finally, we conduct extensive experiments to evaluate the efficacy of TrojanNet in accurately detecting Trojans in QAOA circuits. We also compare the performance of TrojanNet with existing machine learning-based Trojan detection methods for conventional netlists. 

\textbf{Paper Organization:}
\label{subsec:paper_organization}
In Section \ref{sec:background}, we provide background on quantum computing and review the related work. Section \ref{sec:threat_model}
outlines the threat model and the adversarial capabilities. Section \ref{sec:vulnerability analysis} presents the proposed heuristic approach to identify the most vulnerable locations for Trojan insertion and determine the type of Trojan gate with the greatest negative impact on QAOA circuits. Section \ref{sec:Trojannet} focuses on the design and performance analysis of TrojanNet, including details on dataset generation, architecture, training and evaluation procedures, and performance analysis. Section \ref{sec:conclusion} summarizes our findings. 

\section{Background} 
\label{sec:background}
\subsection{Quantum Computation Preliminaries}

\subsubsection{Qubits} Quantum computing  exploits the principles of quantum mechanics to perform computations with extraordinary capabilities compared to classical computers. While classical computers represent information using bits that can be either 0 or 1, quantum computers employ quantum bits or qubits, which can exist in a superposition of both 0 and 1 states simultaneously. The state of a qubit can be represented by a complex vector in a two-dimensional Hilbert space. If we consider a single qubit, its state can be expressed as \cite{nielsen2002quantum}: $|\psi\rangle = \alpha|0\rangle + \beta|1\rangle$, where $|0\rangle$ and $|1\rangle$ are the basis states representing the classical bit values 0 and 1, and $\alpha$ and $\beta$ are complex probability amplitudes that satisfy $|\alpha|^2 + |\beta|^2 = 1$. The probability of measuring the qubit in the state $|0\rangle$ ($|1\rangle$) is $|\alpha|^2$ ($|\beta|^2$).

\subsubsection{Quantum Entanglement}

The power of quantum computing lies in its ability to operate on multiple qubits in entangled states. Entanglement allows qubits to become correlated, even when physically separated, resulting in a highly interconnected system. The state of an entangled system cannot be described independently for each qubit but requires a joint description of all the qubits involved. For instance, a two-qubit entangled state can be represented as:$|\psi\rangle = \alpha|00\rangle + \beta|01\rangle + \gamma|10\rangle + \delta|11\rangle$, where $\alpha$, $\beta$, $\gamma$, and $\delta$ are complex probability amplitudes satisfying the normalization condition $|\alpha|^2 + |\beta|^2 + |\gamma|^2 + |\delta|^2 = 1$.

\subsubsection{Quantum Gates} Quantum gates are employed to manipulate the states of qubits and perform computations. These gates can be represented by unitary matrices that act on the state vector of the qubits. Commonly used quantum gates include the Pauli gates ($X$, $Y$, $Z$), the Hadamard gate ($H$), single-qubit rotation gates ($R_x$, $R_y$ and $R_z$), SWAP and the Controlled-NOT gate ($CNOT$). By applying a sequence of quantum gates to the qubits, quantum algorithms can perform complex calculations and solve problems that are computationally intractable for classical computers.

\subsubsection{Quantum Superposition and Parallelism} Quantum superposition and parallelism are fundamental concepts in quantum computing. Quantum superposition refers to the ability of a quantum system to exist in multiple states simultaneously. In contrast to classical bits, which can be either in a state of 0 or 1, quantum bits or qubits can be in a superposition of 0 and 1. This allows qubits to represent and process multiple states simultaneously, exponentially increasing computational power. Quantum parallelism leverages this property by performing computations on all possible combinations of qubit states in parallel. This enables quantum algorithms to explore multiple solutions simultaneously, offering the potential for significant speedup in solving certain types of problems compared to classical computation.

\subsection{Compilation of Quantum Circuits} The compilation process is a crucial step in quantum computing that transforms high-level quantum circuit representations into executable forms compatible with physical quantum computers. A typical quantum compiler, such as Qiskit, follows several key steps \cite{anis2021qiskit}. Firstly, virtual circuit optimization techniques are applied to reduce gate count and enhance performance. Secondly, multi-qubit gates are decomposed into single- and two-qubit gates to match hardware capabilities. Next, the circuit is mapped onto physical qubits while considering coupling constraints and connectivity limitations. Routing algorithms determine optimal paths for executing the circuit, followed by a translation into basis gates supported by the target device. Finally, physical optimization techniques, such as gate fusion and cancellation, are applied to reduce gate count and execution time further.

\subsection{Quantum Algorithm}
A quantum algorithm is a mathematical representation for performing computations on quantum computers. It leverages the parallelism and superposition of qubits to explore multiple outcomes simultaneously. 

\subsubsection{Quantum Approximate Optimization Algorithm (QAOA)}
\begin{figure}[t]
    \includegraphics[width=0.5\textwidth]{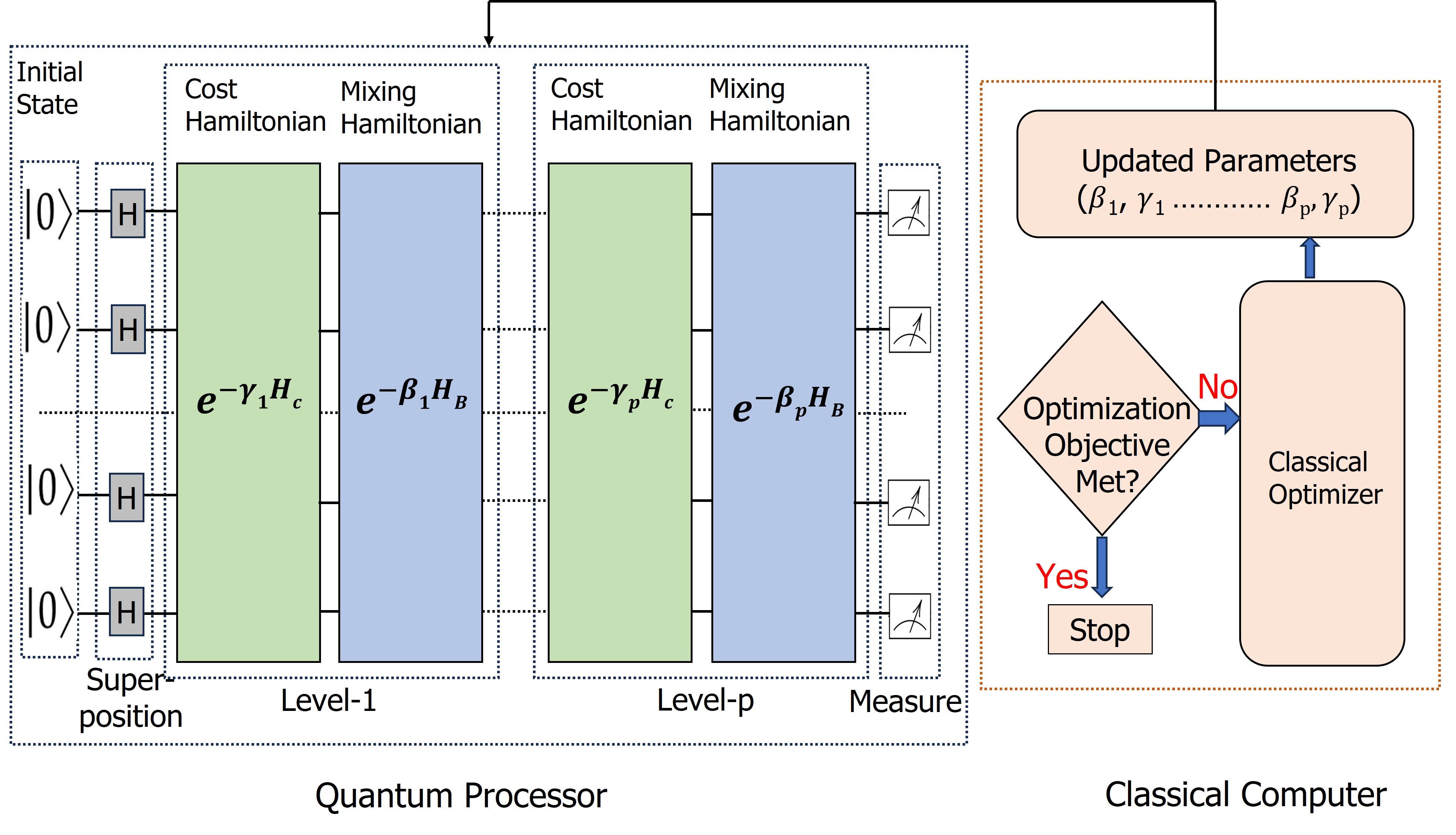}
    \centering
\caption{Schematic of QAOA. The algorithm consists of $p$ layers, where input qubit states undergo alternating applications of the Cost Hamiltonian (CH) and the Mixing Hamiltonian (MH). The final state is measured to obtain the expectation value of the objective function. The classical optimizer finds the optimal parameter values ($\gamma$, $\beta$) that maximize or minimize the cost.}
    \label{fig:QAOA algorithm}
    \vspace{-.31cm}
\end{figure}

QAOA is a powerful hybrid quantum-classical variational algorithm designed to solve combinatorial optimization problems. QAOA operates by iteratively applying a parameterized sequence of quantum gates to the qubits in a quantum circuit. The algorithm consists of alternating layers of the Cost Hamiltonian ($H_C$) and the Mixing Hamiltonian ($H_M$) operations \cite{zhou2020quantum}. The Cost Hamiltonian encodes the objective function of the optimization problem, while the Mixing Hamiltonian introduces quantum fluctuations between different configurations. The structure of the QAOA algorithm is shown in Fig. \ref{fig:QAOA algorithm}.

The algorithm begins with the preparation of an initial state, typically a uniform superposition state. This state is then subjected to a series of alternating layers of quantum gates. Each layer consists of rotations generated by the Cost Hamiltonian, given by the unitary operator $e^{-i\gamma_l H_C}$, followed by rotations induced by the Mixing Hamiltonian, given by the unitary operator $e^{-i\beta_l H_M}$. The rotations introduced by the Cost Hamiltonian allow for the exploration of different configurations, while the Mixing Hamiltonian promotes mixing between different states. The number of layers, denoted as $p$, determines the depth and flexibility of the algorithm in searching the solution space. After the application of the quantum circuit, the final state is measured, yielding a bitstring that represents a potential solution. The classical optimization process, utilizing techniques such as gradient-based methods or heuristics, is then employed to adjust the parameters of the quantum gates to minimize the objective function and find an approximate solution. 

The approximation ratio (AR) is an important metric used to evaluate the performance of QAOA in solving combinatorial optimization problems. It quantifies the quality of the solution obtained by QAOA compared to the optimal solution. AR is defined as the ratio of the objective function value obtained by QAOA to the optimal objective function value:

\[ \text{Approximation ratio (AR)} = \frac{E(\theta)}{E_{\text{opt}}}, \]

where $E_{\text{opt}}$ is the optimal objective function value achieved by classical optimization algorithms or other known methods. The closer the AR is to 1, the better the solution provided by the QAOA algorithm.

\subsubsection{Graph Max-Cut Problem}

The Graph Max-Cut problem is a widely studied combinatorial optimization problem that finds applications in various domains such as network analysis, image processing, and circuit design \cite{crooks2018performance}. Given an undirected graph $G = (V, E)$, where $V$ represents the set of nodes and $E$ represents the set of edges, the objective is to partition the nodes into two disjoint sets $A$ and $B$ such that the number of edges crossing the partition is maximized. In other words, we aim to find a cut $(A, B)$ that maximizes the cut size. Notably, it has been established that achieving an approximation ratio of $AR \geq \frac{16}{17} \approx 0.9412$ for Max-Cut on all graphs is NP-hard \cite{haastad2001some}.

Mathematically, the Graph Max-Cut problem can be formulated as follows. Let $n$ be the number of nodes in the graph. We assign a binary variable $x_i \in \{0, 1\}$ to each node $i$, indicating whether it belongs to set $A$ ($x_i = 1$) or set $B$ ($x_i = 0$). The objective function to be maximized is: $f(x) = \sum_{(i,j) \in E} w_{ij} \cdot (1 - x_i \cdot x_j)$, where $w_{ij}$ is the weight of the edge $(i, j)$. The term $x_i \cdot x_j$ represents the product of the binary variables, indicating whether the nodes $i$ and $j$ belong to different sets. Thus, the expression $(1 - x_i \cdot x_j)$ evaluates to 1 when nodes $i$ and $j$ are in different sets (crossing the cut) and 0 otherwise. The goal is to find the assignment of binary variables $x = (x_1, x_2, \ldots, x_n)$ that maximizes the cut size i.e. $\text{Maximize } \sum_{(i,j) \in E} w_{ij} \cdot (1 - x_i \cdot x_j)$.

\subsubsection{QAOA for Solving Graph Max-Cut Problem}

The QAOA has shown significant promise in solving the Graph Max-Cut problem. QAOA circuits encode the structure of the graph and the corresponding objective function into the qubits' quantum states \cite{crooks2018performance}. By applying parameterized quantum gates in alternating layers, QAOA explores the space of possible cuts and aims to find the one that maximizes the cut size. Mathematically, let's consider a graph with $m$ edges and $n$ vertices. To represent the assignment of vertices to sets A and B, we utilize a bitstring $z = z_1 \ldots z_n$, where $z_i = 0$ corresponds to vertices in set A and $z_i = 1$ corresponds to vertices in set B. The primary objective is to maximize the number of edges that are `cut' by the partition, denoted as $C(z)$. When applying the QAOA to approximate solutions for the Max-Cut problem, we leverage computational basis states $|z_i\rangle$ and define the objective function as follows \cite{huang2023near}:

\[
C(z) = \sum_{\alpha=1}^{m} C_\alpha(z) = \sum_{\text{{edge}}(j,k)}^{m} \frac{1}{2}(1 - \sigma_j^z \sigma_k^z)|z_i\rangle,
\]

Here, $(j, k)$ represents the vertices connected by the $\alpha$-th edge, and $C_\alpha$ takes on a value of 1 only when the $j$-th and $k$-th qubits exhibit different measurement outcomes on the Z basis, indicating separate partitions.

The QAOA circuit designed for Max-Cut typically comprises $p$ layers, each governed by the problem Hamiltonian $H_C = \sum_{\text{{edge}}(j,k)}^{m} \frac{1}{2}(1 - \sigma_z^j \sigma_z^k)$, and a mixing Hamiltonian $H_M = \sum_{i=1}^n \sigma_x^i$, alternatingly applied in each layer. This parameterized circuit can be represented as:

\[
U(\gamma, \beta) = \prod_{l=1}^p e^{-i\beta_l H_M} e^{-i\gamma_l H_C},
\]

Here, $\gamma = (\gamma_1, \ldots, \gamma_p)$ and $\beta = (\beta_1, \ldots, \beta_p)$ denote the variational parameters that determine the rotations and entanglement operations within each layer. To initiate the QAOA, the initial state of the circuit is set to an $n$-qubit uniform superposition state $|+\rangle_n$. After applying the parameterized circuit, the final state $\phi(\gamma, \beta) = U(\gamma, \beta)|+\rangle_n$ is measured, yielding the bitstring $z$ that represents the optimal partition achieving the maximum cut of edges. The training process of the QAOA parameterized circuit aims to guide its evolution from the initial state towards the bitstring $z$ that achieves the maximum partition.

\subsection{Related Work and Their Limitations} 
Several studies have addressed the challenges and vulnerabilities associated with quantum computing. One area of research focuses on the security threats posed by untrusted compilers in quantum circuits \cite{das2023randomized, suresh2021quantum}. They point out potential IP theft issues introduced by unreliable compilers during the optimization process. Another work \cite{saki2021split} presents a split compilation methodology to address the same concern. By splitting the quantum circuit into multiple parts and sending them to a single compiler at different times or to multiple compilers, this methodology provides partial information to adversaries and introduces factorial time reconstruction complexity. Another relevant work \cite{upadhyay2023obfuscating} focuses on obfuscating quantum hybrid-classical algorithms, specifically QAOA, to protect sensitive information encoded in the circuit parameters from untrusted quantum hardware. The proposed edge pruning obfuscation method and split iteration methodology secure the IP while maintaining low overhead costs. Further, \cite{upadhyay2023trustworthy} proposes equal distribution of computations among hardware options and an adaptive heuristic to identify tampered hardware. While previous works have explored quantum security from various angles such as untrusted compiler, the specific issue of Trojan insertion, insertion methods, and detection mechanisms in quantum circuits has not been studied yet to the best of our knowledge. 

In terms of Hardware Trojan (HT) detection in integrated circuits (ICs), several studies using machine learning techniques have been proposed \cite{liakos2019machine}. For instance,\cite{yasaei2022hardware} proposed a golden reference-free HT detection method using Graph Neural Networks (GNNs) for both Register Transfer Level (RTL) and gate-level netlists. The authors leverage a Data Flow Graph (DFG) representation of the hardware design to train the GNN model, enabling the detection of unknown HTs with high recall rates. Another study \cite{zareen2018detecting} explored the use of Artificial Immune Systems (AIS) for detecting RTL Trojans by leveraging the behavior classification of high-level hardware descriptions. Furthermore, \cite{han2019hardware} focused on machine learning-based HT detection at the RTL level, employing the gradient boosting algorithm and a server-client mechanism for timely updates. While previous research has made significant contributions to Trojan detection in traditional ICs, there is a noticeable lack of studies focusing on Trojan detection in quantum circuits, such as QAOA. 
To bridge this gap, we utilize insights from existing machine learning-based research on HT detection and apply them to the unique characteristics and vulnerabilities of QAOA circuits, enabling effective Trojan detection in the quantum computing domain.

\section{Threat model and adversary capability} 
\label{sec:threat_model}
\subsection{Threat Model} 

We assume that the user designs a QAOA circuit to solve the Max-Cut problem and employs untrusted or less-trusted third-party compilers to optimize the depth and gate count. This choice may arise due to the limited availability of trusted compilers that keep pace with the latest optimization advancements. The QAOA circuits sent to untrusted compilers can be subjected to impactful Trojan insertion and/or tampering. 
By introducing Trojans or extra gates in specific locations, adversaries can introduce biases or perturbations that disrupt the optimization process. As a result, the quality and reliability of the obtained solutions are compromised. Once adversaries have strategically inserted Trojans or extra gates into the QAOA circuits, they compile the modified circuits and transmit them to the quantum hardware. The process of compilation often transforms the original circuit into a form that is not easily inspectable or discernible, making it more challenging to identify the presence of Trojans. This inherent difficulty in detecting Trojans within the compiled circuit amplifies the risks associated with compromised optimization outcomes.

\begin{figure*}[t]
    \includegraphics[width=0.9\textwidth]{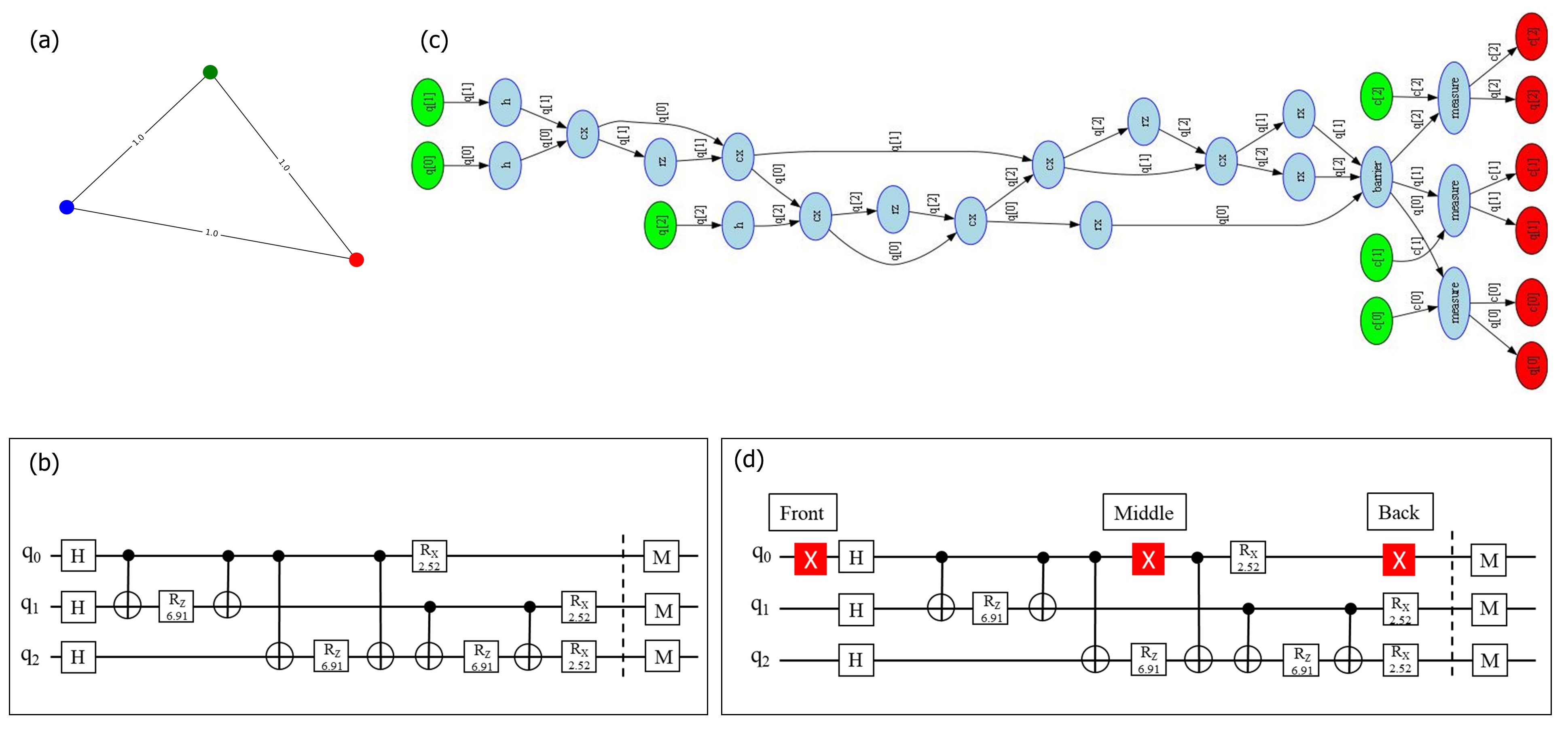}
    \centering
    \caption{Illustration of the case study using a 3-node benchmark graph. (a) The 3-node graph used for the MAx-Cut problem.
(b) The optimized QAOA circuit obtained for solving the MAx-Cut problem. (c) The Directed Acyclic Graph (DAG) representation of the optimized QAOA circuit. (d) Insertion of a Trojan X gate (red gates) at the front, middle, and back positions of the critical path.}
    \label{fig:case_study}
    \vspace{-.31cm}
\end{figure*}

\subsection{Adversary Capability} 

We assume that adversaries possess expertise in the structure and functioning of QAOA circuits. They have a comprehensive understanding of the circuit's components, including the sequence of gates, qubit connectivity, and measurement operations. This knowledge enables them to identify suitable insertion points for extra gates and determine the optimal timing to achieve maximum impact on the circuit's behavior. Moreover, adversaries have access to significant computational resources, allowing them to analyze and manipulate the circuits effectively. With their computational power, they can perform detailed analyses of the circuit's properties, such as entanglement patterns and optimization landscape. This enables them to strategically insert extra gates that introduce biases or modify the entanglement structure, ultimately leading to the generation of suboptimal solutions during the optimization process. 


\section{Trojan Vulnerability Analysis for QAOA Circuits}
\label{sec:vulnerability analysis}

In this section, we present a heuristic approach for conducting vulnerability analysis of QAOA circuits. Using a case study, we determine the most vulnerable locations within the QAOA circuit where the insertion of Trojans (extra gates) can degrade the optimization process the most. We also identify the type of extra gate that has the most negative effect on the performance of the QAOA circuit. Finally, we evaluate the effectiveness of the proposed Trojan-insertion strategy on benchmark QAOA circuits.

\subsection{Overview}
To identify the most vulnerable Trojan insertion locations, we start by selecting a small graph with 3, 4, or 5 nodes. 
Next, we employ the QAOA algorithm to solve the graph Max-Cut problem. By running the QAOA algorithm for a fixed number of iterations, we optimize the circuit's parameters and obtain the corresponding QAOA circuit. This optimized circuit is then utilized to calculate the approximation ratio (AR), which quantifies the quality of the solution obtained. The optimized circuit is then converted into a Directed Acyclic Graph (DAG). By analyzing the DAG, we identify the critical path of the circuit. To determine the most vulnerable insertion location, we introduce an extra gate at the front, middle, and back of both the critical and non-critical paths and compile it using IBM Qiskit software \cite{anis2021qiskit}. Using the compiled Trojan-inserted QAOA circuit, we solve the graph Max-Cut problem with the same number of iterations as the Trojan-free QAOA circuits. We calculate the AR for the Trojan-inserted circuit and analyze the loss in AR compared to the Trojan-free circuit for an equal number of iterations. The location with the highest AR loss indicates the most vulnerable insertion point. Furthermore, we vary the type of Trojan gate inserted at this location and identify the gate type that has the most detrimental effect.

\subsection{Case Study: A 3-Node Graph}

In this subsection, we present a case study to demonstrate the application of the proposed heuristic-based approach using a 3-node benchmark graph, as depicted in Fig. \ref{fig:case_study}(a). All edges of the graph are assigned equal weight. Initially, we employ the QAOA algorithm to solve the Max-Cut problem for this graph and optimize the parameters of the QAOA circuit. We set the number of layers in the QAOA circuit to 1 and utilize the Qiskit for simulations. The optimized QAOA circuit is shown in Fig. \ref{fig:case_study}(b).

\begin{figure}[t]
    \includegraphics[width=0.45\textwidth]{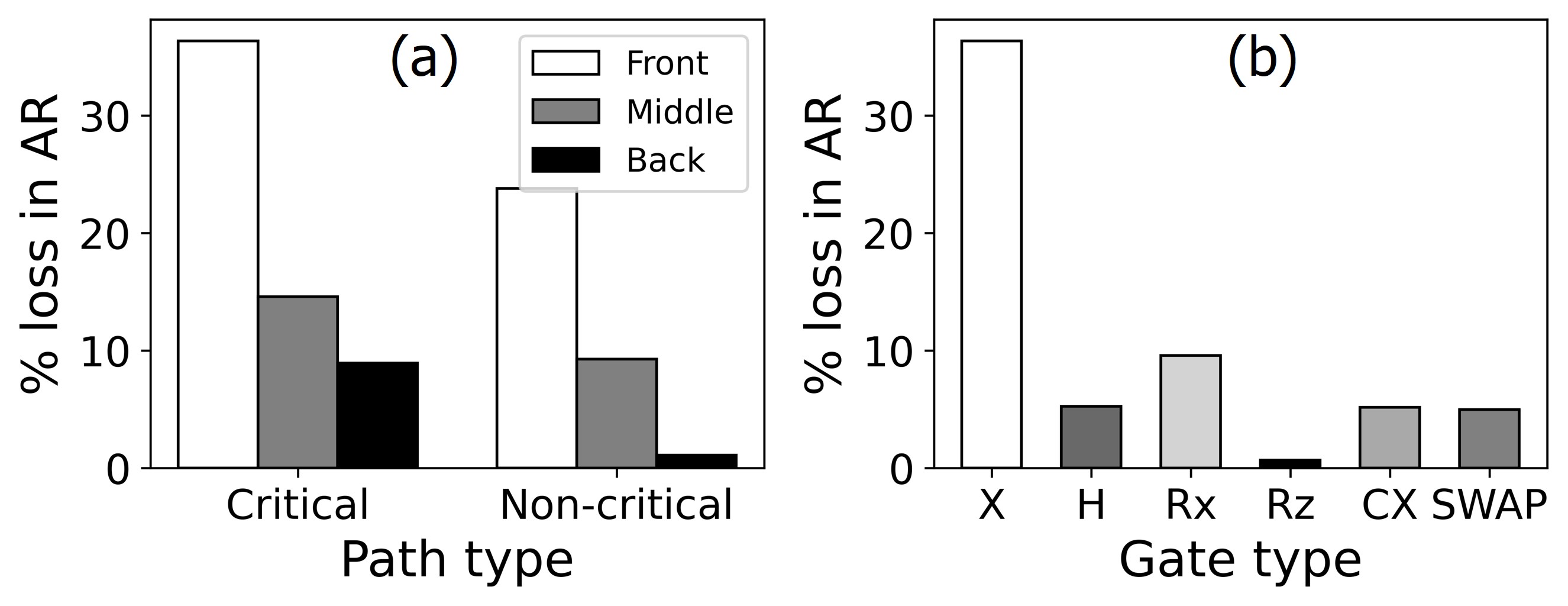}
    \centering
\caption{Impact of Trojan insertion on the AR for the 3-node benchmark graph. (a) Bar chart showing the \%loss in AR due to Trojan insertion at different locations. (b) Comparison of the \%loss in AR for different gate types inserted at the front of the QAOA circuits, including single-qubit gates (X, H, R\textsubscript{x}, R\textsubscript{z}) and two-qubit gates (CX and SWAP).}
    \label{fig:results_3_node}
    \vspace{-.31cm}
\end{figure}

Subsequently, we convert the optimized QAOA circuit into a DAG as illustrated in Fig. \ref{fig:case_study}(c). By analyzing the DAG, we identify the  critical paths within the circuit, namely the paths starting from qubit 0 and qubit 1. To determine the most vulnerable insertion locations, we insert a Trojan X gate at the front, middle, and back positions of both the critical and non-critical paths, as shown in Fig. \ref{fig:case_study}(d). The Trojan-inserted QAOA circuits are then compiled using Qiskit to ensure compatibility with IBM's $qasm\_simulator$ backend. Using the  Trojan-inserted QAOA circuit, we address the same Max-Cut problem for the benchmark graph, employing the same maximum number of iterations (2500) as the Trojan-free circuit. The resulting loss in AR, depicted in the bar chart of Fig. \ref{fig:results_3_node}(a), indicates the impact of Trojan insertion. Interestingly, we observe that the insertion of a Trojan at the front of the critical path leads to a higher loss in AR compared to the non-critical path. Furthermore, we investigate the effect of different gate types inserted at the front of the QAOA circuits, including single-qubit gates (X, H, R\textsubscript{x} with rotation angle 2.52, R\textsubscript{z} with rotation angle 6.91) and two-qubit gates (CX and SWAP). We deliberately chose these rotation angles for the inserted gates to match those of the original R\textsubscript{x} and R\textsubscript{z} gates in the circuits, thereby making the detection of the Trojan more challenging. Strikingly, we find that the loss in AR is notably higher when an X gate is inserted (Fig. \ref{fig:results_3_node}(b)).

Our analysis reveals that the most vulnerable insertion location for Trojans in QAOA circuits is at the front of the critical path. Also, the X gate shows a significant detrimental effect on the circuit's performance. This can be attributed to the disruption caused by the X gate altering the initial state of the qubits. The QAOA algorithm relies on an initial superposition state, typically achieved through H gates, which allows for the exploration of a larger solution space. However, the insertion of an X gate at the beginning disrupts the superposition, limiting the circuit's ability to explore and find optimal solutions. Consequently, the optimization process is hindered, resulting in a higher loss in AR compared to other gate types.

\subsection{Evaluation of Trojan-Insertion Strategy on Benchmark QAOA Circuits}

\begin{figure}[t]
    \includegraphics[width=0.4\textwidth]{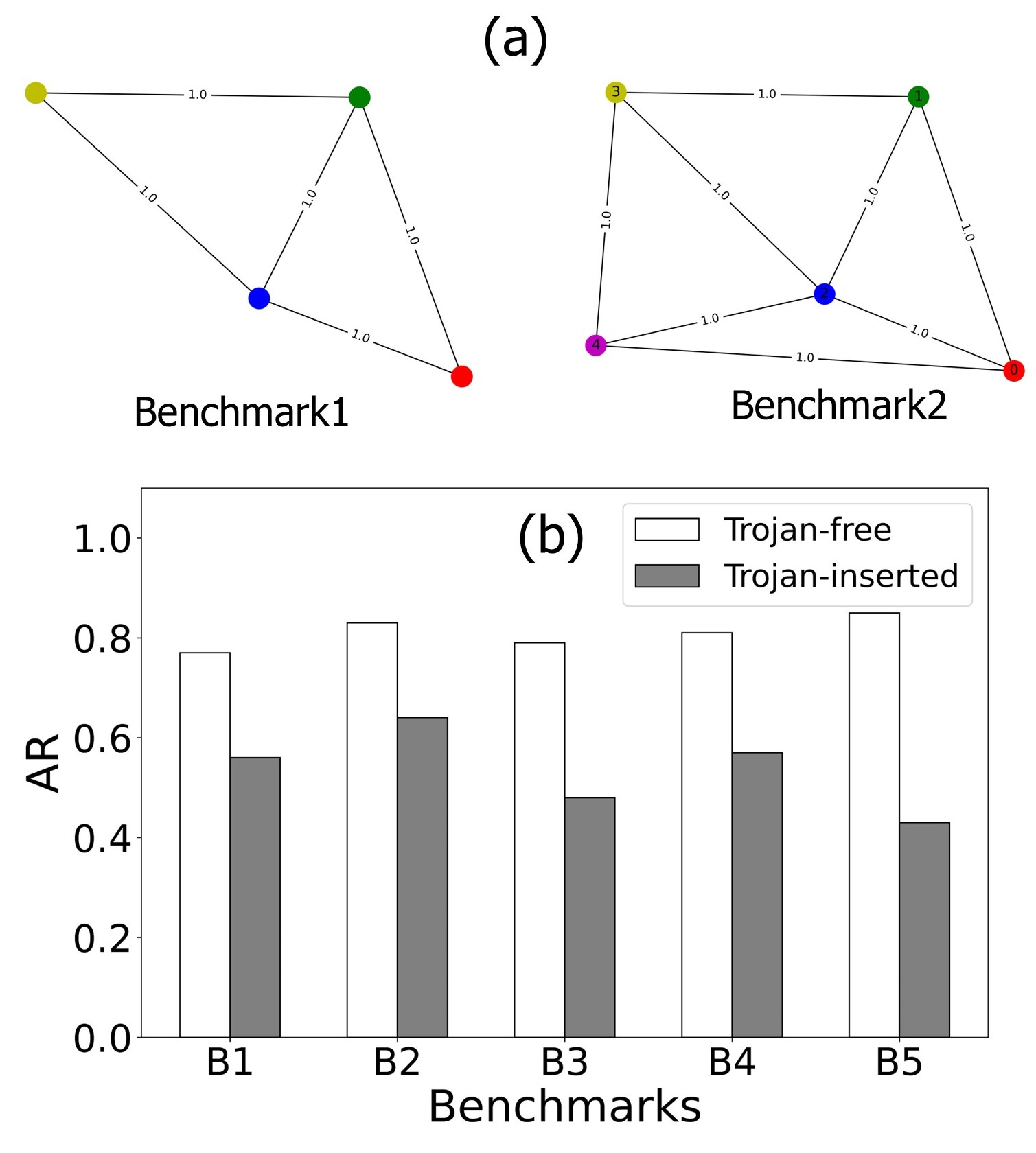}
    \centering
\caption{Evaluation of Trojan-insertion strategy on benchmark QAOA circuits. (a) Sample benchmark graphs for the Max-Cut problem. (b) Approximation ratio comparison between Trojan-free and Trojan-inserted QAOA circuits, showcasing up to 50\% degradation in optimization.}
    \label{fig:results_benchmark}
    \vspace{-.31cm}
\end{figure}

In this subsection, we assess the effectiveness of our identified most impactful Trojan-insertion strategy on a set of five benchmark QAOA circuits. These circuits are specifically optimized for solving the Max-Cut problem of five benchmark graphs consisting of four and five nodes. Two representative benchmark graphs are illustrated in Fig. \ref{fig:results_benchmark}(a). To begin, we utilize the QAOA algorithm and Qiskit to simulate the optimization process on a local machine with an AMD Ryzen 7 4800U CPU operating at 1.80 GHz and 16 GB RAM, running Windows 11 Pro. The number of layers in the optimized QAOA circuits is set to 1. After optimizing the QAOA circuits, we identify the critical path within each circuit by converting them into DAG. Subsequently, we insert a Trojan X gate at the front of the critical path. The Trojan-inserted circuits are then compiled using Qiskit's $qasm\_simulator$ to ensure compatibility and accurate evaluation. Next, we employ the compiled Trojan-inserted circuits to solve the respective graph's Max-Cut problem, using the same number of iterations as the Trojan-free circuits. The resulting ARs for both the Trojan-free and Trojan-inserted benchmark circuits are presented in Fig. \ref{fig:results_benchmark}(b). It demonstrates that the proposed heuristic-based Trojan-insertion strategy can degrade the optimization performance by as much as 50\%. This evaluation provides valuable insights into the vulnerability of QAOA circuits to Trojan insertion and underscores the importance of developing robust defenses against potential attacks.

 \section{Design and Performance Analysis of TrojanNet}
\label{sec:Trojannet}
 
In this section, we delve into the design and performance analysis of TrojanNet. We discuss the dataset generation process, the architecture of TrojanNet, the training and evaluation procedures, and present the performance results and evaluation metrics of the model.
\subsection{Dataset Generation}
To begin with, we generate a total of 813 unique graphs with 3, 4, and 5 nodes using Qiskit. These graphs are designed to meet the requirements for the Max-Cut problem, ensuring that each node is connected to at least one edge. Using the QAOA algorithm, we solve the Max-Cut problem for each generated graph and optimize the parameters of the QAOA circuit. The optimized circuits, referred to as `Trojan-free QAOA circuits', are then compiled using Qiskit's $qasm\_simulator$ backend. 
Next, we insert a trojan X gate at the front of the critical path within each non-compiled Trojan-free QAOA circuit by applying the strategy which led to the highest AR loss. These Trojan-inserted circuits are then compiled using the $qasm\_simulator$. Thus, our dataset comprises 813 Trojan-free and 813 Trojan-inserted compiled QAOA circuits.

We further create 11 additional datasets that incorporate variations in Trojan gate types (X/ H/ R\textsubscript{x}/ CNOT), number of gates (1 or 2), insertion location (front/ middle) and compiler backend ($qasm\_simulator$/ $FakeManilaV2$). While not all of these trojan insertion strategies show a significant loss in AR, this approach allows us to assess the efficiency and effectiveness of TrojanNet across diverse scenarios.

\subsection{Architecture of TrojanNet}
\begin{figure}[t]
    \includegraphics[width=0.45\textwidth]{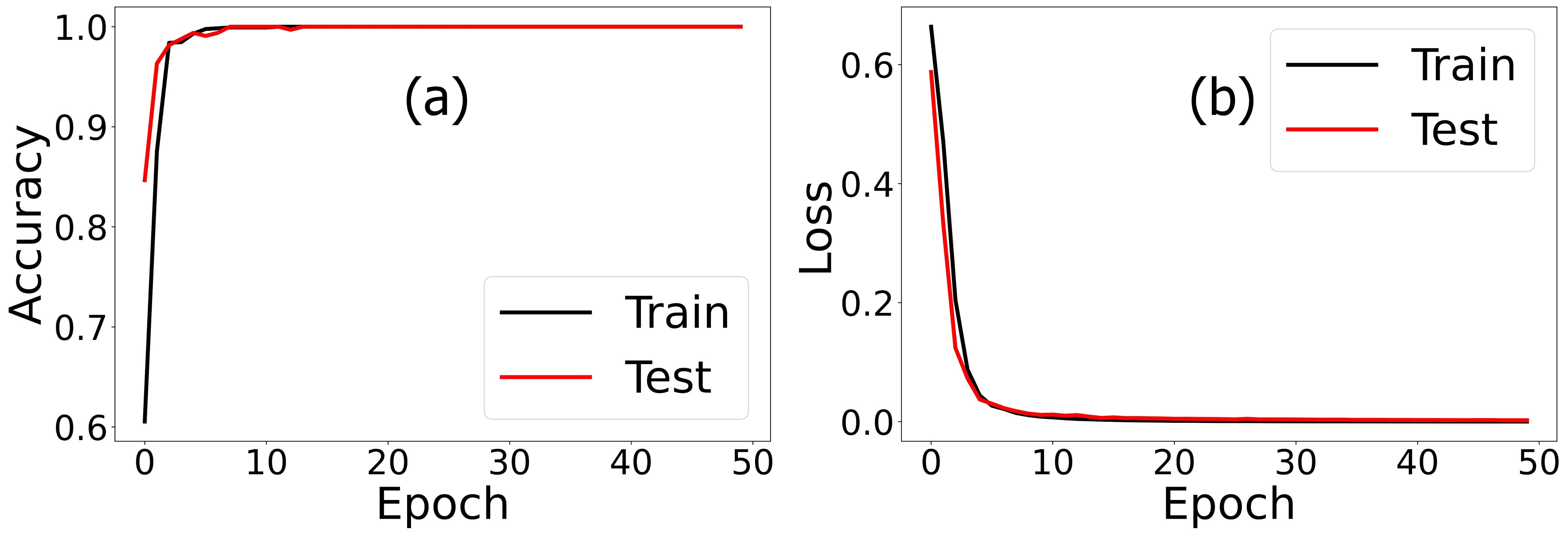}
    \centering
\caption{(a) Accuracy and (b) Loss vs Epoch plots demonstrating TrojanNet's training progress and performance.}
    \label{fig:results_cnn}
    \vspace{-.31cm}
\end{figure}

TrojanNet is implemented as a CNN model using the TensorFlow framework, specifically the Keras API \cite{geron2022hands}. It is specifically designed to effectively differentiate between Trojan-free and Trojan-inserted compiled QAOA circuits by leveraging the inherent characteristics of the circuit representations. TrojanNet involves several key components and techniques to capture subtle patterns and features within the circuits. The CNN model consists of multiple layers that operate on the 2D representations of the QAOA circuits obtained by converting the Quantum Assembly Language (QASM) files to unitary matrices using Qiskit's \textit{Operator} method. These unitary matrices represent the quantum gates in the circuit, which serve as the input data. The unitary matrices are padded or truncated to a common size to ensure uniformity. 

TrojanNet starts with a Convolutional layer with 32 filters and a kernel size of 3x3, which performs feature extraction by convolving over the input unitary matrices. The Rectified Linear Unit (ReLU) activation function is applied to introduce non-linearity. A MaxPooling layer with a pool size of 2x2 follows the Convolutional layer, reducing the spatial dimensions and capturing important features. The pooling operation aids in preserving the most relevant information while reducing computational complexity. The output of the MaxPooling layer is flattened into a 1D vector, which is then fed into a fully connected Dense layer with 64 units. This layer further extracts higher-level features and applies non-linearity through the ReLU activation function. Finally, the output layer consists of two units with a softmax activation function, representing the two classes: Trojan-free and Trojan-inserted. The softmax activation function normalizes the outputs and provides probabilities for each class.

\subsection{Training and Evaluation} 
The dataset is split 80\%-20\% into training-testing set for model evaluation. The training methodology involves optimizing the model's weights and biases through an iterative process. The Adam optimizer is employed, which adaptively adjusts the learning rate and performs efficient parameter updates. During training, the TrojanNet model learns to extract meaningful features and patterns from the circuit representations. The training dataset is split into training and validation sets, enabling the model to generalize well to unseen data. The model is evaluated on the validation dataset to monitor its performance and prevent overfitting during training. The categorical cross-entropy loss function is used to measure the discrepancy between the predicted and actual labels. The model aims to minimize this loss by adjusting its parameters.  

To assess the performance of the TrojanNet model, various evaluation metrics are utilized. While accuracy is a commonly used metric, it may not capture the model's robustness against different scenarios. Therefore, additional metrics, such as precision, recall, and F1-score, are considered, providing insights into the model's performance in correctly identifying both Trojan-free and Trojan-inserted circuits. Precision measures the proportion of correctly identified Trojan-inserted circuits among the predicted positives. Recall, also known as sensitivity, calculates the proportion of Trojan-inserted circuits correctly identified among the actual positives. The F1-score combines precision and recall, offering a balanced measure of the model's performance.

\subsection{Results and Analysis}

\begin{table}[t]
\centering
\caption{Performance of TrojanNet for Different Datasets}
\label{tab:table1}
\resizebox{\columnwidth}{!}{%
\begin{tabular}{|c|c|c|c|c|c|c|c|}
\hline
\rotatebox[origin=c]{90}{Backend} & \rotatebox[origin=c]{90}{Location} & \rotatebox[origin=c]{90}{Gate Type} & \rotatebox[origin=c]{90}{\# of Gate} & \rotatebox[origin=c]{90}{Accuracy} & \rotatebox[origin=c]{90}{Precision} & \rotatebox[origin=c]{90}{Recall} & \rotatebox[origin=c]{90}{F1-Score} \\ \hline
\multirow{6}{*}{\rotatebox[origin=c]{90}{Qasm}}         & Front   & X         & 1            & 100.00\% & 100.00\%  & 100.00\% & 100.00\% \\ \cline{2-8} 
                               & Front    & H         & 1            & 100.00\% & 100.00\%  & 100.00\% & 100.00\% \\ \cline{2-8} 
                               & Front    & R\textsubscript{x}        & 1            & 98.77\%  & 98.11\%   & 98.11\%  & 98.11\%  \\ \cline{2-8} 
                               & Front    & CX        & 1            & 100.00\% & 100.00\%  & 99.37\%  & 99.68\%  \\ \cline{2-8} 
                               & Middle   & R\textsubscript{x}        & 1            & 98.47\%  & 98.70\%   & 95.60\%  & 97.12\%  \\ \cline{2-8} 
                               & Middle   & R\textsubscript{x}        & 2            & 98.77\%  & 98.14\%   & 99.37\%  & 98.75\%  \\ \hline
\multirow{6}{*}{\rotatebox[origin=c]{90}{FakeManilaV2}} & Front   & X         & 1            & 99.03\%  & 98.72\%   & 99.35\%  & 99.04\%  \\ \cline{2-8} 
                               & Front    & H         & 1            & 99.01\%  & 98.69\%   & 99.23\%  & 99.05\%  \\ \cline{2-8} 
                               & Front    & R\textsubscript{x}        & 1            & 97.49\%  & 97.30\%   & 97.30\%  & 97.30\%  \\ \cline{2-8} 
                               & Front    & CX        & 1            & 99.35\%  & 99.35\%   & 98.89\%  & 99.03\%  \\ \cline{2-8} 
                               & Middle   & R\textsubscript{x}        & 1            & 97.10\%  & 97.23\%   & 94.64\%  & 96.45\%  \\ \cline{2-8} 
                               & Middle   & R\textsubscript{x}        & 2            & 97.56\%  & 97.93\%   & 98.49\%  & 97.86\%  \\ \hline
\end{tabular}%
}
\end{table}

\begin{table*}[t]
\centering
\caption{Comparing the performance of TrojanNet with existing ML-based Hardware Trojan Detection Technique}

\label{tab:table2}
\begin{tabular}{|l|l|l|l|l|}
\hline
Type of  Trojan                  & Technique   Catageroy- Method                 & Precision & Recall  & Reference \\ \hline
\multirow{6}{*}{Hardware Trojan in conventional circuits} & ML - Graph neural network on RTL graph        & 92\%      & 97\%    & \cite{yasaei2022hardware} \\ \cline{2-5} 
                                 & ML - Graph neural network on netlist graph    & 91\%      & 84\%    & \cite{yasaei2022hardware}  \\ \cline{2-5} 
                                 & ML - Artificial immune system                 & 87\%      & 85\%    & \cite{zareen2018detecting} \\ \cline{2-5} 
                                 & ML - Gradient boosting algorithm              & NA        & 100\%   & \cite{han2019hardware}  \\ \cline{2-5} 
                                 & ML - Multi-layer neural networks              & NA        & 90\%    &  \cite{hasegawa2017hardware}         \\ \cline{2-5} 
                                 & CA - Socio-network analysis                   & 98\%      & 98\%    &  \cite{islam2019socio}        \\ \hline
Trojan in quantum circuit        & ML-Convolutional Neural Network on QASM files & 98.68\%   & 98.36\% & Our work  \\ \hline
\end{tabular}%
\end{table*}

The trained TrojanNet model is evaluated using the generated datasets. Performance results for a specific dataset, created with the most impactful trojan insertion strategy (X gate at the front, $qasm\_simulator$ backend), are shown in Fig. \ref{fig:results_cnn}. The model was trained for 50 epochs, with accuracy and loss monitored (Fig. \ref{fig:results_cnn}(a)). Training and validation accuracies gradually increased, reaching 100\% validation accuracy at epoch 10. The model loss decreased with increasing epochs, indicating convergence (Fig. \ref{fig:results_cnn}(b)). Additional evaluation metrics, including precision, recall, and F1-score, were calculated. For this dataset, precision, recall, and F1-score all achieved a perfect score of 100\%, demonstrating the model's ability to accurately detect Trojan-inserted circuits. 

Table \ref{tab:table1} further presents the evaluation results of TrojanNet on the remaining datasets. As can be seen, the accuracy of TrojanNet in detecting and classifying Trojan-inserted QAOA circuits ranges from 97.10\% to 100.00\%, with an average accuracy of 98.80\%. Noticeably, when using the $FakeManilaV2$ backend, the accuracy is slightly lower compared to the $qasm\_simulator$ backend. This discrepancy can be attributed to the increased number of gates in the compiled circuits due to qubit mapping and swap operations, which introduce additional complexity. However, overall performance remains impressive, as indicated by the high precision, recall, and F1-score values across all datasets.



\subsection{Comparison with Classical Hardware Trojan Detection Technique}

Table \ref{tab:table2} compares the performance of TrojanNet with existing ML-based techniques in terms of precision and recall metrics for detecting HTs in conventional circuits. Among the classical HT detection techniques, GNN applied to RTL and netlist graphs achieved high precision and recall values \cite{yasaei2022hardware}. An artificial immune system-based approach and a gradient boosting algorithm also demonstrated notable performance \cite{zareen2018detecting, han2019hardware}. The proposed CNN-based TrojanNet achieves higher precision and recall metrics in detecting Trojans in quantum circuits, outperforming these existing techniques.

\section{Conclusion}
\label{sec:conclusion}

We investigate the vulnerability of QAOA circuits to Trojan insertion during compilation by untrusted third parties and develop TrojanNet for detection and classification. 
We identified the most vulnerable insertion location for Trojans in QAOA circuits to be at the front of the critical path, with an X gate having the most detrimental effect. The evaluation of our Trojan-insertion strategy on benchmark QAOA circuits revealed up to a 50\% loss in the approximation ratio. TrojanNet demonstrated remarkable accuracy, reaching 98.80\% in accurately differentiating between Trojan-free and Trojan-inserted QAOA circuits. The dataset and the TrojanNet tool will be released to the public to further the research on quantum security. 

\section*{Acknowledgements}
This work is supported in parts by NSF (CNS-1722557, CNS-2129675, CCF-2210963, CCF-1718474, OIA-2040667, DGE-1723687, DGE-1821766, and DGE-2113839), Intel’s gift and seed grants from Penn State ICDS and Huck Institute of the Life Sciences. 

\bibliography{Ref}

\end{document}